\documentclass[sigconf, natbib=false, nonacm]{acmart}
\usepackage[datamodel=acmdatamodel, style=acmnumeric, backend=biber]{biblatex}
\usepackage{amsmath, amsfonts}

\usepackage{graphicx}
\usepackage{subcaption}

\usepackage[nameinlink]{cleveref}

\usepackage[inline]{enumitem}
\setlist*[itemize]{labelindent=10pt, itemindent=0pt, leftmargin=*}

\usepackage{booktabs}
\usepackage{multirow}
\usepackage{tabularx}

\usepackage{pgfplots}
\pgfplotsset{compat=1.18}
\usepgfplotslibrary{statistics}

\usepackage{csquotes}
\usepackage{textcomp}

\usepackage{balance}

\setcopyright{rightsretained}
\copyrightyear{2024}

\acmDOI{xx.xxx/xxx_x}
\acmISBN{xxx-x-xxxx-xxxx-x/xx/xx}

\acmYear{2024}
\acmConference[SAC'24]{ACM SAC Conference}{April 8 –April 12, 2024}{Avila, Spain}

\begin{document}

\title[Reduced Simulations for High-Energy Physics]{Reduced Simulations for High-Energy Physics, a Middle Ground for Data-Driven Physics Research}

\author{Uraz Odyurt}
\orcid{0000-0003-1094-0234}
\affiliation{%
	\institution{High-Energy Physics, Radboud University}
	\city{Nijmegen}
	\country{The Netherlands}}
\affiliation{%
	\institution{National Institute for Subatomic Physics (Nikhef)}
	\city{Amsterdam}
	\country{The Netherlands}}
\email{uodyurt@nikhef.nl}

\author{Stephen Nicholas Swatman}
\orcid{0000-0002-3747-3229}
\affiliation{%
	\institution{Informatics Institute, University of Amsterdam}
	\city{Amsterdam}
	\country{The Netherlands}}
\affiliation{%
	\institution{European Organisation for Nuclear Research (CERN)}
	\city{Geneva}
	\country{Switzerland}}
\email{s.n.swatman@uva.nl}

\author{Ana-Lucia Varbanescu}
\orcid{0000-0002-4932-1900}
\affiliation{%
	\institution{CAES, University of Twente}
	\city{Enschede}
	\country{The Netherlands}}
\affiliation{%
	\institution{Informatics Institute, University of Amsterdam}
	\city{Amsterdam}
	\country{The Netherlands}}
\email{a.l.varbanescu@utwente.nl}

\author{Sascha Caron}
\orcid{0000-0003-2941-2829}
\affiliation{%
	\institution{High-Energy Physics, Radboud University}
	\city{Nijmegen}
	\country{The Netherlands}}
\affiliation{%
	\institution{National Institute for Subatomic Physics (Nikhef)}
	\city{Amsterdam}
	\country{The Netherlands}}
\email{scaron@nikhef.nl}

\renewcommand{\shortauthors}{U. Odyurt et al.}

\begin{abstract}
Subatomic particle track reconstruction (tracking) is a vital task in High-Energy Physics experiments. Tracking is exceptionally computationally challenging and fielded solutions, relying on traditional algorithms, do not scale linearly. Machine Learning (ML) assisted solutions are a promising answer. We argue that a complexity-reduced problem description and the data representing it, will facilitate the solution exploration workflow. We provide the REDuced VIrtual Detector (REDVID) as a complexity-reduced detector model and particle collision event simulator combo. REDVID is intended as a simulation-in-the-loop, to both generate synthetic data efficiently and to simplify the challenge of ML model design. The fully parametric nature of our tool, with regards to system-level configuration, while in contrast to physics-accurate simulations, allows for the generation of simplified data for research and education, at different levels. Resulting from the reduced complexity, we showcase the computational efficiency of REDVID by providing the computational cost figures for a multitude of simulation benchmarks. As a simulation and a generative tool for ML-assisted solution design, REDVID is highly flexible, reusable and open-source. Reference data sets generated with REDVID are publicly available. Data generated using REDVID has enabled rapid development of multiple novel ML model designs, which is currently ongoing.
\end{abstract}

%
%

\ccsdesc[500]{Applied computing~Physics}
\ccsdesc[500]{Computing methodologies~Simulation types and techniques}
\ccsdesc[500]{Computing methodologies~Simulation tools}

\keywords{Reduced-order modelling, Simulation, High-energy physics, Synthetic data}

\maketitle


\section{Introduction}
\label{sec:introduction}
In many computational sciences, the adoption of ML-assisted solutions can lead to serious gains in computational efficiency and data processing capacity, resulting from algorithmic advantages intrinsic to ML. Computational efficiency can also be achieved by paving the way for the utilisation of dedicated hardware, i.e., GPUs, FPGAs and purpose-built accelerators. ML algorithms are highly compatible with the use of such specialised hardware. In this work, we explore the use of ML-assisted techniques in high-energy physics. 

\emph{The process of ML-assisted solution design is an explorative and data-demanding endeavour}. One of the effective approaches to achieve a suitable design is Design-Space Exploration (DSE). However, DSE suffers from the fact that direct exploration for solutions addressing complex problems consume high amounts of time and energy. Complex problems involve many parameters, contributing to a space with many dimensions, which in turn deems the exploration expensive. As a result, there is often a need for simplification of the problem domain, i.e., search-space reduction, to facilitate the initial steps within this explorative process.

Generative elements are often needed as part of the explorative process, to enable synthetic data generation in large quantities, at will. Furthermore, designing and training models with better rigour requires total control over all aspects of data generation. Providing sufficient control and on demand ability to synthesise data that is representative of corner cases contributes to achieving effective models. Such corner cases seldom/disproportionally appear in real-world data or highly accurate, i.e., \emph{physics-accurate}, simulation data.

\paragraph*{Use-case}
Our focus is a major use-case from the field of High-Energy Physics (HEP), \emph{the critical task of subatomic particle track reconstruction (tracking)}, which is present in data processing for experiments performed at the Large Hadron Collider (LHC). Detectors such as ATLAS, record interaction data of subatomic particles with detector sensors, allowing physicists to reconstruct particle trajectories through tracking algorithms and to gain knowledge on how subatomic particles behave. The current tracking solutions relies largely on traditional, computationally expensive statistical algorithms, with Kalman filtering as their most demanding block. Even with constant efforts channelled into better parallelisation schemes for these algorithms, the data consumption capability is rather limited. The challenge will be even greater with the upcoming High-Luminosity LHC upgrade~\cite{Apollinari:2014:HLLP}, given its increased data volume generation and experiment frequency.

Although physics-accurate simulators, such as Geant4~\cite{Agostinelli:2003:GST}, are readily available, applying such levels of accuracy to generative elements comes at a hefty computational cost. Accordingly, these simulators are not suitable for frequent \emph{timely} executions and constant data generation, as required for DSE iterations. As such, we propose an exploration methodology that can be much faster, through the informed simplification of the design-space for the ML-assisted solution. Our methodology is specifically being considered for the tracking use-case. To this end, we have \emph{designed and implemented} the \emph{REDuced VIrtual Detector (REDVID)}, to both simplify the problem at hand and act as an efficient tool for frequent simulations and synthetic data generation. While our tool is not a fully physics-accurate one, it does respect the high-level relations present in subatomic particle collision events and detector interactions. REDVID is fully (re)configurable, allowing definition of experiments through varying detector models, while preserving the \emph{cascading effects} of every change.

Considering possible complexity reduction strategies, the spectrum varies from physics-accurate data manipulations, e.g., dimensionality/granularity reduction, to omitting the scenario interactions beforehand. A strategy solely based on data reduction will fail to preserve the behavioural integrity of the system, as it will fail to propagate cascading effects resulting from reductions. Even simplified examples such as the TrackML data~\cite{Amrouche:2020:TMLC} are too complex.

\paragraph*{Contribution}
We provide REDVID, an experiment-independent, fully (re)configurable, and complexity-reduced simulation framework for HEP~\cite{Odyurt:2023:REDVIDSITE}. Simulations consist of complexity-reduced detector models, alongside a particle collision event simulator with reduced behavioural-space. REDVID is intended as a simulation-in-the-loop for ML model design workflows, providing:
\begin{itemize}
	\item ML model design - Problem simplification facilitates ML solution design, as opposed to real-world use-case definitions, which are often too complex to negotiate directly.
	\item Parametric flexibility - The model generator is capable of spawning detectors based on reconfigurable geometries.
	\item Computational efficiency - Behavioural-space reductions directly improve event simulation and processing times.
\end{itemize}
Our other contributions include:
\begin{itemize}
	\item Supporting pedagogical tasks in higher education by presenting complex interactions from HEP experiments through simplified and understandable data.
	\item Publishing open reference data sets, which are of independent interest for physicists and data scientists alike~\cite{Odyurt:2023:DATASET1, Odyurt:2024:DATASET2}.
\end{itemize} 

\paragraph*{Outline}
\Cref{sec:background} provides the background on HEP experiments and similar simulators. In \Cref{sec:design}, we provide the design details considered for REDVID. Notable implementation techniques are elaborated in \Cref{sec:implementation}. Data set related results are given in \Cref{sec:results}, followed by \Cref{sec:related_work,sec:conclusion}, covering the relevant literature and our conclusions, respectively.

\section{Background and motivation}
\label{sec:background}
In this section we elaborate the premise of HEP experiments, as well as the role of simulation in these, to get familiar with the context of our use-case.

\subsection{HEP experiments}
When talking about \emph{HEP experiments}, we refer to high-energy particle collision events. Two types of collision experiments are performed at LHC: proton-proton and ion-ion collisions. Protons are extracted from hydrogen atoms, while ions are actually heavy lead ions. Beams of particles are sent down the beam pipe in opposing directions and made to collide at four specific spots. These four spots are the residing points of the four major detectors installed at LHC, namely, ALICE~\cite{Collaboration:2008:ALICE}, ATLAS~\cite{Collaboration:2008:ATLAS}, CMS~\cite{Collaboration:2008:CMS} and LHCb~\cite{Collaboration:2008:LHCb}.

Take the ATLAS detector for instance. The role played by ATLAS in the study of fundamental particles and their interactions, rely on two main tasks, \emph{tracking} and \emph{calorimetry}. Through tracking, i.e., particle track reconstruction, the momentum, $p$, of a particle can be calculated, while the energy, $E$, is calculated through calorimetry. Having the momentum and the energy for a given particle, its mass, $m$, can be determined, following the \emph{energy-momentum relation} expressed as,
\[
E^2 = (mc^2)^2 + (pc)^2 \texttt{.}
\]

In the above equation, $c$ represents the speed of light and is a constant. The mass measurement allows the study of the properties for known particles, as well as potentially discovering new unknown ones. As such, it is fair to state that \emph{particle track reconstruction is one of the major tasks in high-energy physics}.

\subsection{Role of simulation in HEP}
Simulation allows for, amongst others, the validation and training of particle track reconstruction algorithms. Two distinguished stages are considered for HEP event simulations, i.e., \emph{physics event generation} and \emph{detector response simulation}~\cite{Edmonds:2008:FATS}. Event generation as the first stage, involves the simulation of particle collision events, encompassing the processes involved in the initial proton-proton or ion-ion interactions. Event generation is governed by intricate sets of physical rules and is performed by software packages such as Herwig~\cite{Corcella:2001:EGHE} and Pythia~\cite{Sjöstrand:2006:PYTH}, i.e., physics-accurate simulations.

Detector response simulation, the second stage, integrates the movement of the particles generated by the first stage through a detector geometry, simulating the decay of unstable particles, the interactions between particles and matter, electromagnetic effects, and further physical processes such as hadronisation. Common event simulators providing such functionality include Geant4~\cite{Agostinelli:2003:GST}, FLUKA~\cite{Böhlen:2014:FCDC} and MCNP~\cite{Forster:2004:MCNP}. In accelerator physics applications, event simulators are used to simulate the interactions between particles and sensitive surfaces in an experiment, as well as with so-called passive material, such as support beams. Interactions with sensitive surfaces may undergo an additional \emph{digitisation} step, simulating the digital signals that can be read out of the experiment. Considering the example of ATLAS, three data generating simulators are notable, namely, Geant4, FATRAS~\cite{Edmonds:2008:FATS} and ATLFAST~\cite{Richter-Was:1998:AFSP}.

Following the Monte Carlo simulation approach, FATRAS has been designed to be a fast simulator. It is capable of trajectory building based on a simplified reconstruction geometry and does provide support for material effects, as well as particle decay. FATRAS also generates hit data.

ATLFAST follows a different approach towards trajectory simulation and doesn't generate hit data, making it unsuitable for tracking studies. ATLFAST relies on hard-coded smearing functions based on statistics from full simulations. These functions are dependent on particle types, momentum ranges and vertex radii. Such details are specific to the design elements of the virtual detector geometry. A change in the design will require finding new functions.

\begin{figure}[htbp]
	\centering
	\includegraphics[width=0.7\linewidth]{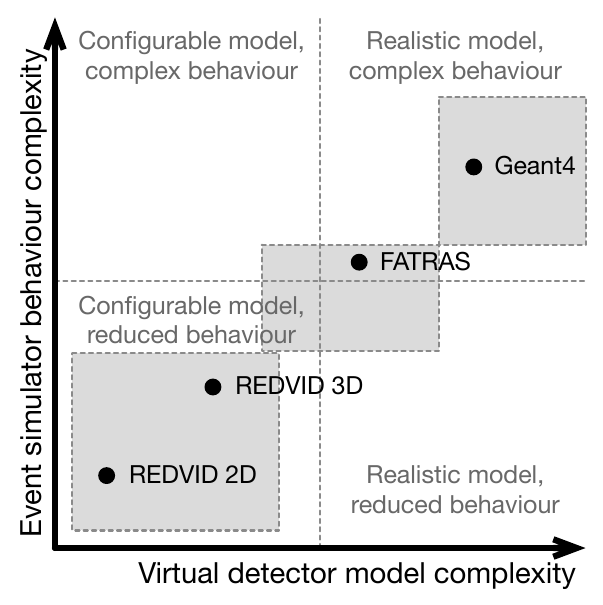}
	\caption{Simulation complexity spectrum is shown from the most simplistic to the most realistic, with high complexity rates for both model and simulator. Depending on the enabled features, different simulators are capable of providing different levels of complexity, depicted as grey areas. ATLFAST is not included for lack of hit data generation. Note that the figure does not cover data reduction strategies, which are not relevant to changes in model or simulator complexity.}
	\label{fig:simulation_spectrum}
\end{figure}

REDVID fills the gap for a reconfigurable framework that is suitable for first-phase solution exploration and design. This is due to the deliberate reduction in complexity, for both the generated data and the problem description, while keeping the high-level causal relations in place. REDVID is end-to-end parametric, i.e., all the generated data is built upon the detector geometry and randomised particle trajectories, both reconfigurable. REDVID has been developed in Python, making its integration with Python-based ML design workflows seamless. \Cref{fig:simulation_spectrum} plots REDVID's positioning versus other well-known tools, as we consider it.

\section{Simulation application and design}
\label{sec:design}
The underlying question here is what is a good strategy for designing and training a capable and rigorous ML model to predict the behaviour of a (complex) real system? The higher the complexity of the system and its associated data, the harder it is to arrive at an efficient ML model design solving the task. Generally speaking, complex tasks require larger models, which in turn require more training and more data. For our HEP use-case, the system is already complex; and when considering the upcoming High-Luminosity LHC upgrade~\cite{Apollinari:2014:HLLP}, this complexity will increase even further. As such, when looking for an ML-assisted solution for HEP tracking, we need to efficiently explore a large set of options, and will consequently require lots of data. 

Addressing complex real-world tasks directly will require synthesising close to real-world data, which can be performed by high-accuracy simulations. High-accuracy simulations in general, and physics-accurate simulations in particular, are extremely expensive computational tasks. Having such tools as part of an exploration workflow, e.g., a ML model design workflow, triggering frequent executions of the simulation with altered configuration, will inevitably turn into a serious challenge. Even if there are accommodating hardware resources available, algorithmic limitations will turn software tools into workflow bottlenecks. Yet another notable drawback is the high cost of energy when running frequent computationally expensive tasks. To alleviate this massive challenge, it is highly beneficial, and perhaps necessary, to not only design reduced models and simulators\footnote{A model and a simulator go hand in hand to form a simulation.}, but \emph{to provide parametric (re)configurability to support automated exploration}.

\begin{figure*}[htbp]
	\centering
	\includegraphics[width=0.6\linewidth]{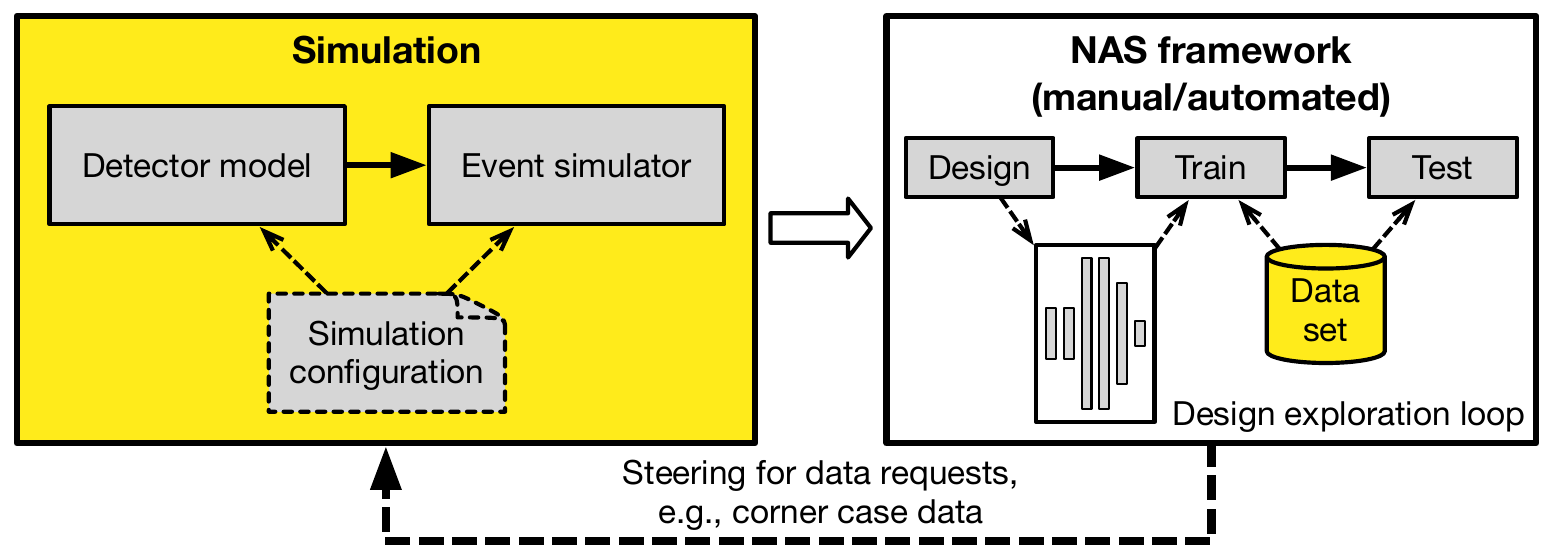}
	\caption{An overview of a reduced simulation as part of a ML model design workflow, e.g., a Neural Architecture Search (NAS), by providing the data set. This paper focuses on the area with the yellow fill, covered by our simulation tool, REDVID.}
	\label{fig:simulation_application}
\end{figure*}

However, the initial testing of new ML-assisted solutions, i.e., ML model designs, does not require the ground truth, which physics-accurate simulations are designed to produce. Instead, we argue that a cost-effective and reduced simulation, preserving the behavioural relations of the complex system (proton-proton/ion-ion collision event experiments), can be better and more effectively integrated in ML model design workflows, as shown in \Cref{fig:simulation_application}.

\subsection{Reduction approach}
Simulations of complex systems include virtual models that mimic the behaviour of the system under scrutiny. On the one hand, the amount of detail included in the model, as a virtual representation of the complex system, will directly affect the approximation level of the simulation. On the other hand, the extent of behaviour considered by a simulator while executing the model will determine the overall achieved complexity. \emph{Having a validly approximate representation is achieved through the reduction of the behavioural-space to a minimal subset, best encapsulating the complex system}. Both model complexity and simulator complexity can be targets of such a reduction. The first and foremost effect of an approximate simulation is better computational efficiency. Note that there can be many such approximations, depending on the intended balance between computational efficiency and behavioural approximation level. The other advantage, especially when it comes to ML model design processes, is facilitation of an effective model design by providing a middle ground that has a lower complexity and can be used for better understanding of the challenge and testing of the early designs, before addressing the full real-world case.

Solution exploration/design in general, and solutions based on ML models in particular, has always benefited from methodical simplification of the problem at hand. For a system operating over a broad behavioural-space, such a simplification is often manifested by means of high-level modelling. Both actual experiments and physics-accurate simulations for our use-case, i.e., proton-proton/ion-ion collision events inside a detector such as ATLAS, are immensely complex. Removing (some of) the physics-accurate constraints results in major behavioural-space reductions. This applies to both the detector model and the behaviour affecting the event simulator. While moving away from physics-accuracy, our aim has been to conserve logical, mathematical and geometrical relations, which would provide the basis for a flexible parameterisation. Preserving relations between interacting elements of a system preserves occurrence of \emph{cascading effects} when the system is being steered through reconfiguration. For instance, a change in the structural definition of the detector model will affect the recorded hit points during the event simulation. It must be noted that we have intentionally avoided the time dimension complexities. Accordingly, a list of major reductions that we have considered follows.

\paragraph*{Simplified detector geometry}
The real detector has a complex geometry, with many small sensors acting collectively to record particle interactions. For instance, a barrel sub-detector type is built using many smaller modules, ultimately forming a cylindrical shape. There are also supporting subsystems, e.g., for cooling, which occupy parts of the detector space. We have considered much simpler elements for the geometry of our virtual detector model, consisting of elements with disk or cylinder shapes, ultimately arriving at a Reduced-Order Model (ROM).

\paragraph*{Particle types}
The particle type plays a major role in its traversal path through the detector. In fact, as stated in \Cref{sec:background}, one of the major applications of track reconstruction is to assign the particle type. Currently, we do not consider explicit particle types in our event simulator. The track type variation however, could be seen as a consequence of differing particle types.

\paragraph*{Simplified tracks}
In the real detector, tracks follow an arc of helix like path and not an exact one. This path is not the case for all particles and the charged characteristic of a particle of interest is a defining factor. Currently, we consider particles traversing a linear (straight), helical uniform, or helical expanding paths. The linear case indirectly suggests that either the single particle type we consider is one with no charge, or alternatively, we do not consider a magnetic field, which is present in the real detector. Helical tracks could be seen as the effect of a magnetic field on charged particles.

\paragraph*{Collision points}
The real experiments involve multiple collisions happening almost at the same time. The collision points will not be at the same spot. Even the collision of interest that is intended for track reconstruction will not perfectly align to the detector's origin point. The neighbouring collisions will also pollute the detector readings with particles associated to them. We consider a single event at the origin for linear tracks and a non-aligning one for helical tracks, i.e., origin smearing.

\paragraph*{Hit coordinates smearing}
When it comes to instrumentation noise, there is no well-defined grand complication present. The amount of noise in real experiments depends on the characteristics of the sensors and material. We introduce noise in our hit calculations and hit coordinate parameters by drawing random samples from a Gaussian distribution. We also consider the noise standard deviation as proportional to the variable range. Like the rest of REDVID's features, the noise ratio can be adjusted by the user.

\subsection{Detector model}
\begin{figure}[htbp]
	\centering
	\includegraphics[width=0.9\linewidth]{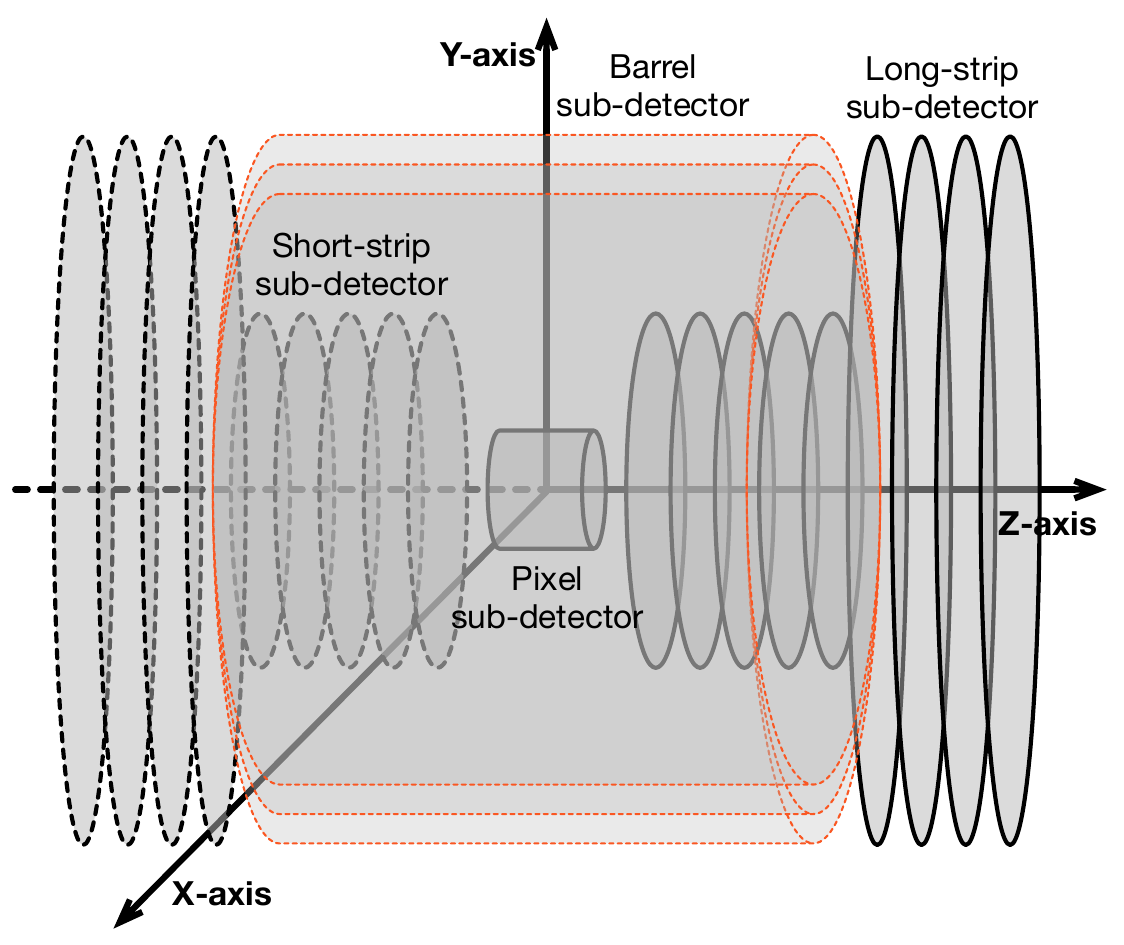}
	\caption{The fully parametric detector geometry, allowing for inclusion/exclusion of different sub-detector types, with full control over sub-layer counts, sizes and placements.}
	\label{fig:detector_geometry}
\end{figure}

At its core, a detector model is comprised of the geometric definitions of the included elements, shapes, sizes, and placements in  space. Although we can support a variety of detector geometries, the overall structure, especially for our experimental results, is based on the ATLAS detector. Accordingly, there are four sub-detector types, \emph{Pixel}, \emph{Short-strip}, \emph{Long-strip} and \emph{Barrel}. The pixel and the barrel types have cylindrical shapes with the pixel being a filled cylinder, while the barrel being a cylinder shell with open caps. These are not hard requirements, as the geometry is fully parametric, and differing definitions can be opted for, e.g., a pixel as a cylinder shell. The long-strip and the short-strip types are primarily intended as flat disks, but can be defined as having a thickness, rendering them as cylinders. Sub-detector types can be selectively present or absent. \Cref{fig:detector_geometry} depicts a representative variation of the detector geometry involving the aforementioned elements.

Structurally speaking, in a real-world detector, like ATLAS, the internals of short-strip and long-strip sub-detector types are different. We on the other hand, reduce such complexities to placement location and size, i.e., distance from the origin and sub-detector disk radius. Note that our geometric model does support disk thickness, which basically would turn disks into shallow cylinders. However, we have considered flat disks for our experiments.

\subsection{Particle collision event simulation}
As mentioned above, one of the simplifications for our complexity reduction approach is to consider a single collision per event, aligning exactly to the origin point of the detector geometry for the linear case. However, the list of complexities, even without the polluting effects of multiple collisions, is extensive. Particles travelling through the detector matter could lead to secondary collisions, resulting in drastic changes in their trajectory. Such secondary collisions could also lead to the release of particles not originating from the collision event itself. These will show up as tracks with unusual starting points within the detector space, rather distant from the collision point. Some particles could also come to a halt, which would be seen as abruptly terminating tracks.

Such physics-accurate behaviour of particles interacting with the present matter in detectors is not considered for our simulator. It must be noted that the generation of tracks originating far away from the origin and prematurely terminating tracks, can be emulated in our simulator in a randomised fashion.

\section{Implementation}
\label{sec:implementation}
Though our detector generator and event simulation modules support both two-dimensional (2D) and three-dimensional (3D) spaces, we will focus on the implementation details relevant to the three-dimensional case. Let us simply mention that the main difference between the two would be the presence of circles and cylinders for 2D and 3D spaces, respectively. One can consider the rather simplistic 2D space as a form of sanity check set-up for initial testing of techniques and methodologies of ML-assisted solution workflows. REDVID is open source~\cite{Odyurt:2023:REDVIDREPO} and has been developed in Python.

\subsection{Modules}
Considering the tasks at hand, detector spawning and event simulation, our software can be divided into three main logical modules:
\begin{itemize}
	\item Detector generator - To spawn a detector based on the provided geometric specifics and configuration.
	\item Event simulator - To execute experiments involving many events, following the experiment configuration, e.g., hit probability, number of tracks (fixed/variable), track randomisation protocol, etc.
	\item Reporting - To collect the expected output, i.e., the generated data set, as well as automated report generation on important configuration and a statistical overview of the data set.
\end{itemize}

\begin{figure*}[htbp]
	\centering
	\includegraphics[width=0.8\linewidth]{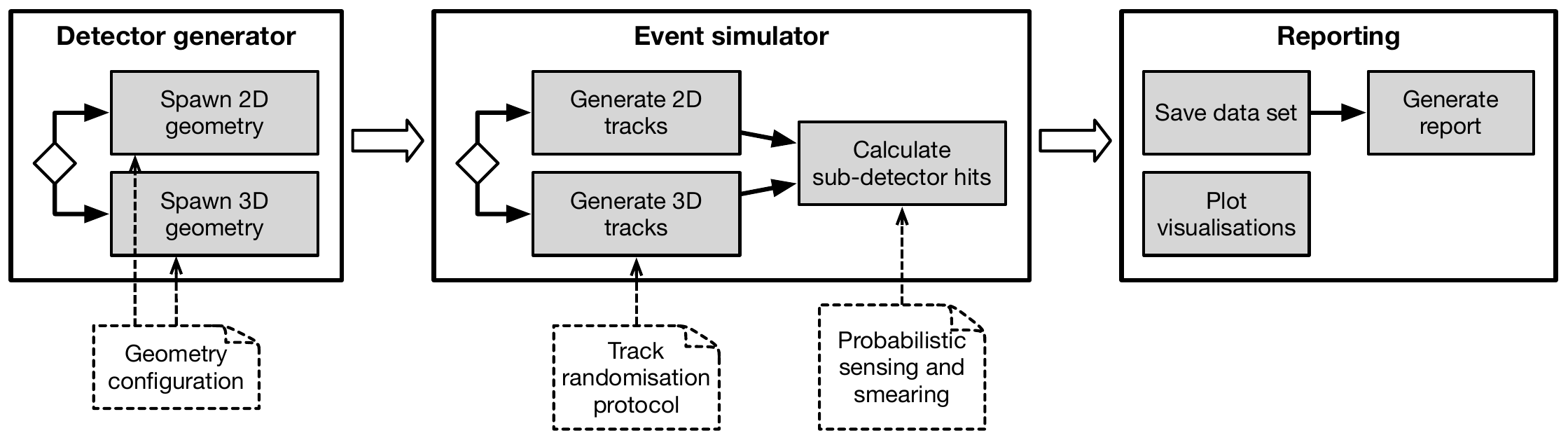}
	\caption{An overview of the REDVID modules, including a detector model generator, an event simulator, generating randomised tracks and calculating sub-detector hit points based on tracks and geometric data, as well as different reporting elements.}
	\label{fig:implementation_modules}
\end{figure*}

An overview diagram of the modules is depicted in \Cref{fig:implementation_modules}. The current implementation considers the sequential execution of modules in the order given above. However, one can easily generate detectors without simulating events, or simulate events with previously generated detectors, or even calculate hits based on previously generated tracks. Such input/output capability will allow our software to interact with other commonly utilised tools. The main configuration parameter defining the execution path within our tool is the \texttt{detector\_type}, which can be 2D or 3D.

\subsection{Coordinate systems}
\begin{figure}[htbp]
	\centering
	\includegraphics[width=0.6\linewidth]{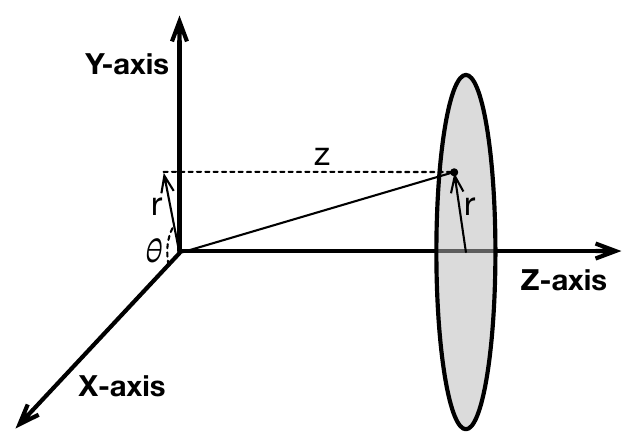}
	\caption{Basic definition and parameters of the cylindrical coordinate system, radial distance, azimuthal, height ($r$, $\theta$, $z$), which is the basis of our geometric structures.}
	\label{fig:cylindrical_coordinate_system}
\end{figure}

For the case of the 3D space, we have opted for the cylindrical coordinate system to represent all elements, i.e., sub-detectors, tracks and hits. The cylindrical coordinate system, depicted in \Cref{fig:cylindrical_coordinate_system}, is a convenient choice, as we are considering the Z-axis as the beam pipe in LHC experiments and all geometric shapes defined within a detector, whether disks or cylinders, are actually of the type cylinder. The three parameters to define any point in the cylindrical coordinate system are the radial distance from the Z-axis, the azimuthal angle between the X-axis and the radius, and the height of the point from the XY-plane, i.e., $r$, $\theta$ and $z$, respectively.

Note that in terms of the orientation of the coordinate system, we consider the Z-axis to be horizontal. With the assumption of the beam pipe's alignment along the Z-axis, this is the most convenient orientation for defining different geometric elements.

In this coordinate system, hit points can be precisely defined given the tuple $(r_{hit}, \theta_{hit}, z_{hit})$. Geometric shapes can also be defined with boundaries for $r_{sd}$ and $z_{sd}$, e.g., a disk will have fixed $z_{sd}$, unbounded $\theta$ and bounded $r_{sd}$. Here $sd$ stands for sub-detector. Our software does support partial disks, i.e., a disk with a hole in the middle, which can be considered when the beam pipe is expected to be part of the geometry. Disks with thickness (cylinders) will have a small boundary for the parameter $z_{sd}$. As previously explained, short-strip and long-strip sub-detector types are defined as disks. For the pixel type, as it is a filled cylinder, both $r_{sd}$ and $z_{sd}$ will be bounded. When it comes to the barrel type, as it is a cylinder shell, there will be a fixed $r_{sd}$ with bounded $z_{sd}$.

To implement linear tracks and to define them in the cylindrical coordinate system, both a direction vector and a point, $P_0$, that the track (line) goes through are needed. The direction vector, $V_d$, is considered as a vector from the origin, landing on a point in space, represented with a tuple $(r_d, \theta_d, z_d)$. The direction vector is randomised and then normalised for the $z$ parameter, meaning that the direction vector will either have $z_d = 1$ or $z_d = -1$. The boundaries of this randomisation depend on the track randomisation protocol, explained in the next section. If we consider all linear tracks as starting from the detector origin, the point $(0, 0, 0)$ is considered on the track. However, this is rarely the case. The resulting parametric form of a track (line) is,
\begin{align*}
	r & = r_0 + t \cdot r_d \texttt{,} \\
	\theta & = \theta_0 + \theta_d \texttt{,} \\
	z & = z_0 + t \cdot z_d \texttt{,}
\end{align*}
with $(r, \theta, z)$ representing a point on the track, $(r_0, \theta_0, z_0)$ as the origin point, $\langle r_d, \theta_d, z_d \rangle$ as the direction vector, and $t$ as free variable.

Following a similar approach, the parametric form for non-linear, i.e., helical, track definitions is,
\begin{align*}
	r &= r_0 + a \cdot t \texttt{,} \\
	\theta &= \theta_0 + d \cdot t \texttt{,} \\
	z &= z_0 + b \cdot t \texttt{,}
\end{align*}
with $(r, \theta, z)$ representing a point on the track and $(r_0, \theta_0, z_0)$ as the origin point, while $a$, $d$ and $b$ represent radial, azimuthal and pitch coefficients, respectively.

Regarding both linear and helical tracks, our software supports origin smearing, i.e., the starting point of helical tracks is in a randomised vicinity of the point $(0, 0, 0)$.

\subsection{Track randomisation protocols}
As seen in \Cref{fig:implementation_modules}, the track randomisation step directly affects sub-detector hit calculation and is totally dependent on the randomisation protocol indicated in the configuration. Focusing on the implementation for the 3D space, different track randomisation protocols can be considered. We list four base protocols and five combination protocols, mixing the characteristics of base protocols:

\paragraph{Protocol 1 - Last layer hit guarantee}
Hits are guaranteed to occur on the farthest layer of every sub-detector type, which means the farthest layer of every sub-detector type is the randomisation domain for the landing points of tracks. A hit guarantee on the last layer will also guarantee hits on the previous layers for that sub-detector type. This protocol is designed to maximise the number of hits per sub-detector type within the data set.

In principle, our implementation applies \emph{Protocol 1} per each available sub-detector type and randomly selects from the total generated track pool. Since for instance, if a track lands on the last layer of strip sub-detector types, it might not necessarily result in hit points on barrel layers.

\paragraph{Protocol 2 - Spherically uniform distribution}
To have a more uniform distribution of randomised tracks, without imposing any geometric conditions, is to have the track end points land on a sphere. Note that tracks do not have actual end points as these are unbounded lines.

\paragraph{Protocol 3 - Conical jet simulation}
Tracks are randomised in distinct subsets, bundled in a close vicinity within a narrow cone, representing a jet(s). This protocol on its own may not be a sensible choice and it would work best in combination with other protocols.

\paragraph{Protocol 4 - Beam pipe concentration}
Tracks will have a higher concentration around the beam pipe, i.e., higher track generation probability as the radius gets smaller.

\paragraph{Protocols 1 and 3}
While still landing on the last sub-detector layer, there are distinct subsets of tracks bundled in a close vicinity as jets. In other words, jets will be mixed with regular tracks.

\paragraph{Protocols 1 and 4}
While still landing on the last sub-detector layer, the tracks landing on the short-strip and the long-strip sub-detector types will have a higher concentration around the beam pipe, i.e., higher track generation probability as the radius gets smaller for these sub-detector types. Tracks landing on the barrel sub-detector type will not be affected.

\paragraph{Protocols 2 and 3}
While still having uniformly distributed tracks landing on a sphere, there will be uniformly distributed distinct subsets of tracks bundled in a close vicinity as jets.

\paragraph{Protocols 3 and 4}
The tracks will have a higher concentration around the beam pipe, i.e., higher track generation probability as the radius gets smaller. There will be jet formation also with higher probability of occurring around the beam pipe.

\paragraph{Protocols 1, 3 and 4}
This combination is the same as the previous, protocols 3 and 4, with the additional condition that the tracks are guaranteed to land on the last layer per sub-detector type.

Note that for our data generation we have only considered protocol 1 to increase recorded hit points for all tracks and to have hit points for all sub-detector types. Needless to say, additional track randomisation protocols focusing on specific corner cases, can be easily defined and added to the tool.

To implement protocol 1, i.e., to guarantee that tracks land on the last layer of a sub-detector type, we consider the coordinate domain of the last layer as the randomisation domain for track direction vectors. Thus, before normalisation, all randomised $V_d$ will land on the last layer.

\begin{figure}[htbp]
	\centering
	\includegraphics[width=0.8\linewidth]{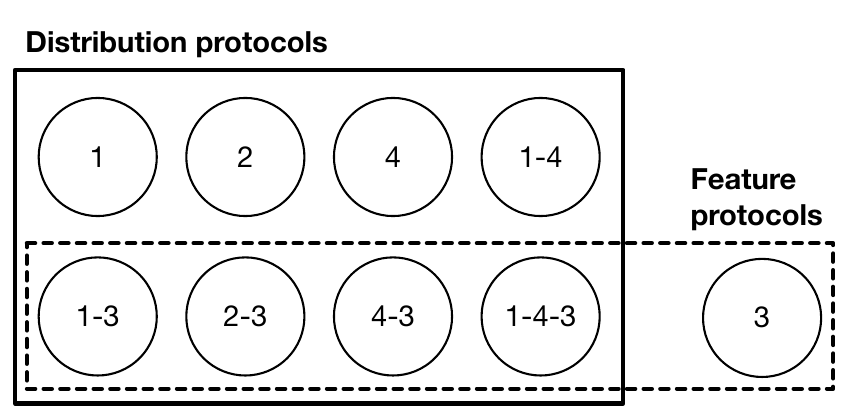}
	\caption{Visualising how different base distribution and feature protocols can be combined to achieve more complex track randomisation behaviour.}
	\label{fig:protocol_combinations}
\end{figure}

As it can be deduced from the above protocol descriptions, not every combination is allowed, as some of the base protocols are mutually exclusive. For instance, protocols 1 and 2 cannot be applied at the same time, as it is self-evident that a spherical uniform distribution and a last layer hit guarantee cannot be true at the same time. Accordingly, we can consider the base protocols within two main categories, \emph{distribution protocols}, affecting how tracks are distributed in space, and \emph{feature protocols}, defining special forms of localised distribution. Currently, protocol 3 is the only feature protocol defined. While feature protocols can be combined with any distribution protocol, most distribution protocols are mutually exclusive. A combination of two or more base distribution protocols will also lead to another, more specific, distribution protocol, e.g., protocols 1 and 4. The diagram in \Cref{fig:protocol_combinations} provides a visual overview of different protocol combinations.

\subsection{Hit point calculation}
Regarding hit point coordinates, i.e., $(r_{hit}, \theta_{hit}, z_{hit})$, depending on the sub-detector shape, we are dealing with either a fixed $z_{sd}$ or a fixed $r_{sd}$, for disks and barrels, respectively. Here, we consider the disks as being flat and to have no thickness, while the barrels consist only of cylinder shells, with no thickness. Shapes with thickness are supported, for which the techniques involved will be similar.

Considering the set of track equations, we are to calculate the free variable $t$ at the sub-detector layer of interest. This specific $t$ is denoted as $t_{sd}$, i.e., $t$ at sub-detector. For hit coordinates at disks,
\begin{align*}
	z_{hit} & = z_{sd} \texttt{,} \\
	\theta_{hit} & = \theta_d \texttt{,} \\
	t_{sd} & = \frac{z_{sd}}{z_d} = \frac{z_{sd}}{1} \texttt{,} \\
	& \Rightarrow t_{sd} = z_{sd} \texttt{,} \\
	r_{hit} & = t_{sd} \cdot r_d = z_{sd} \cdot r_d \texttt{.}
\end{align*}

Note that in the above calculation $z_d$ and $z_{sd}$ must have matching signs, rendering $t_{sd} > 0$. In other words, tracks extruding towards the positive or the negative side of the Z-axis can hit sub-detector layers present at the positive or the negative side of the Z-axis, respectively. We also know that $z_d \neq 0$.

\begin{figure}[htbp]
	\centering
	\begin{subfigure}{0.9\linewidth}
    	\includegraphics[width=\linewidth]{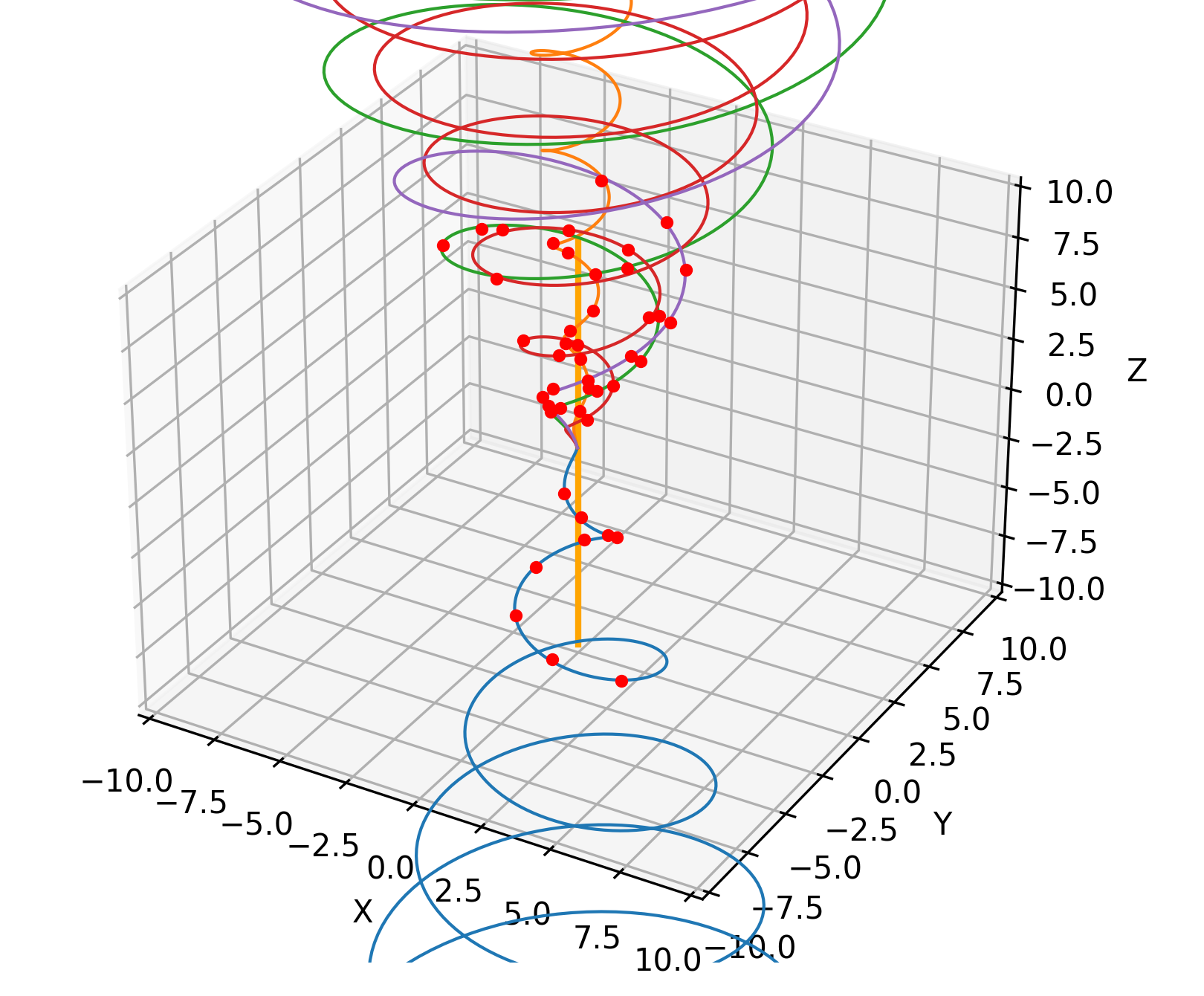}
    	\caption{The full view of this event}
    	\label{fig:example_event_full}
    \end{subfigure}
    \qquad
	\begin{subfigure}{0.9\linewidth}
    	\includegraphics[width=\linewidth]{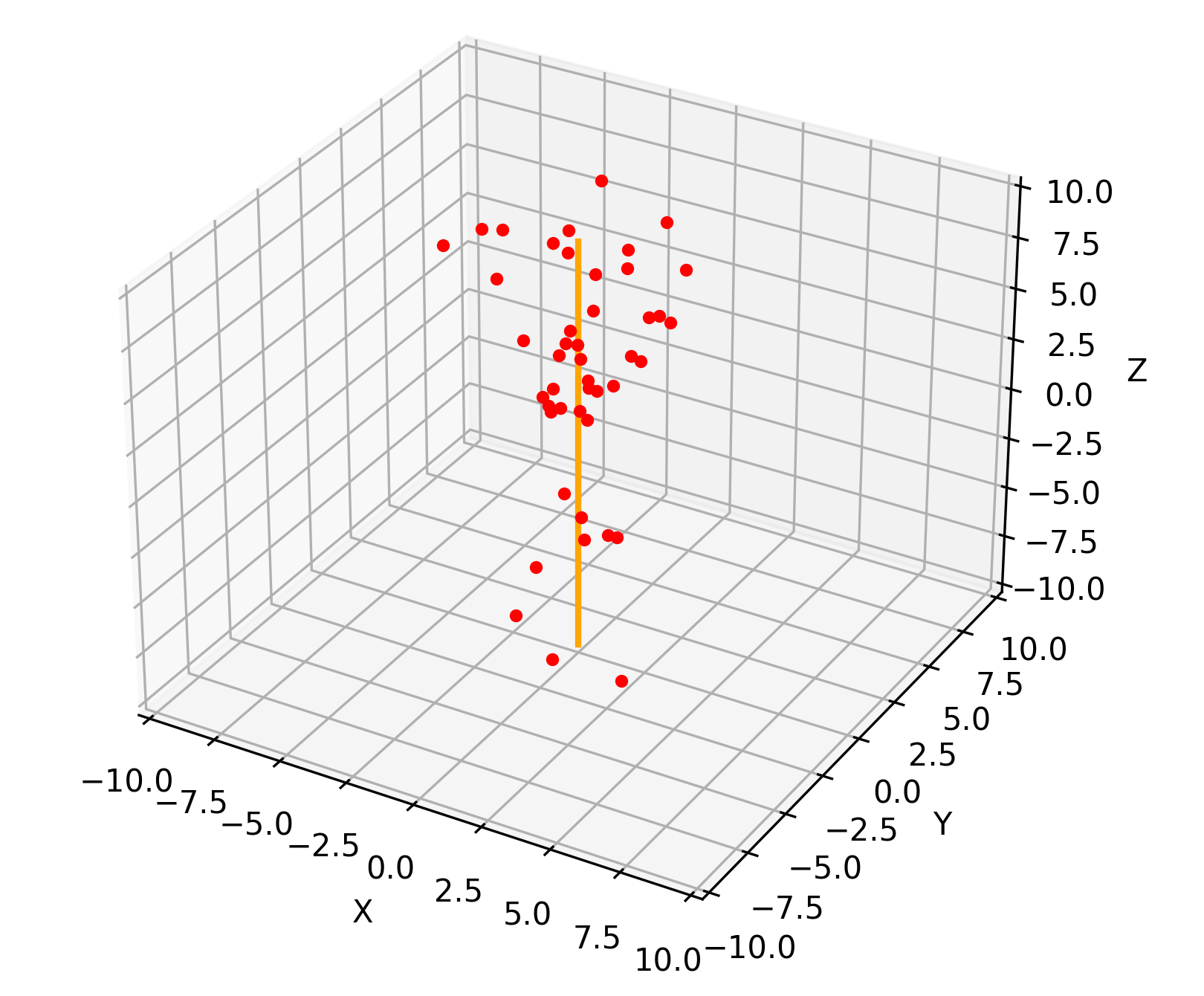}
    	\caption{The hits view of this event}
    	\label{fig:example_event_his}
    \end{subfigure}
    \caption{An example event with five tracks}
	\label{fig:example_event}
\end{figure}

A similar calculation considering the $r_{sd}$ as fixed will result in the hit coordinates for a barrel sub-detector layer, which we will not repeat here. General approach towards calculation of hits resulting from helical tracks follows the same principles, which we will not repeat here. \Cref{fig:example_event} depicts a simple event with five tracks, including separate views of the full event (\Cref{fig:example_event_full}) and calculated hits (\Cref{fig:example_event_his}), for demonstration purposes. Note that the detector orientation is vertical.

\subsection{Available configuration}
We have pointed out a few important configuration options in \Cref{fig:implementation_modules}, i.e., geometry options as a whole, track randomisation and options related to sensing and smearing probabilities when recording hits. Looking at the available options in further detail, REDVID is highly \emph{(re)configurable}.

\paragraph*{3D geometry options}
It is possible to set the detector ID\footnote{The detector ID can be set to auto-generate as well.}, the coordinates for the origin and the centre of each element, the presence of different sub-detector types, thick or flat structure, span over the radius and the $z$ parameters including inner and outer radii, sub-layer counts per sub-detector type, and the distance between consecutive sub-layers per sub-detector type. Spawned detector geometries can be saved for future use.

\paragraph*{Experiment options}
It is possible to set the experiment name, track type, hit coordinate smearing (noise generation), event count, fixed or variable track count with minimum and maximum bounds for the latter, track randomisation protocol, and hit occurrence probability.

\paragraph*{Generic execution options}
A complete folder structure is constructed, only requiring an anchor path to be configured. Additionally, the execution handler can operate in different modes, i.e., \emph{Sequential}, \emph{Parallel}, \emph{Batch Sequential} and \emph{Batch Parallel}. While parallelisation improves speed, automated detection of large simulation jobs, followed by automated batching, is aimed at memory-efficiency. Batch Parallel mode incorporates batching and intra-batch parallelisation at the same time, improving both aspects.

\paragraph*{Non-determinism invoking options}
Though it is desirable to reduce the behavioural-space, as we have done extensively, it is of utmost importance not to arrive at a deterministic simulation. In REDVID, we invoke non-determinism in the simulated behaviour by allowing different randomisations per event, i.e., mandatory randomisation of track parameters, optional track count randomisation, optional introduction of smearing for hit point parameters (noise), and optional hit point occurrence (recording) probability.

Resulting from the modular design, different intermediate data input/output points can be arranged, allowing REDVID to interact with other available tooling. For instance, track data generated by external Monte Carlo event generators can be used alongside a spawned detector geometry to calculate hit points. Needless to say, the input data has to be in a format compatible with REDVID.

\section{Data set generation}
\label{sec:results}
We have considered a number of workloads consisting of both detector spawning and event simulation tasks. We have followed simulation recipes with 10\,000 events and varying track counts of $[1, 10\,000]$ per event for each experiment, for both linear and helical tracks. These recipes are listed below. Hit recording is performed with smearing enabled and the detector geometry is the same for all recipes. These generated data sets for linear and helical tracks are intended as reference for physicists and data scientists alike and are publicly accessible over Zenodo open repository~\cite{Odyurt:2023:DATASET1, Odyurt:2024:DATASET2}.
\begin{itemize}
	\item \texttt{exp-10k-events-1-tracks} - 10\,000 events, 1 track per event, hit coordinate smearing enabled
	\item \texttt{exp-10k-events-10-tracks} - 10\,000 events, 10 tracks per event, hit coordinate smearing enabled
	\item \texttt{exp-10k-events-100-tracks} - 10\,000 events, 100 tracks per event, hit coordinate smearing enabled
	\item \texttt{exp-10k-events-1k-tracks} - 10\,000 events, 1\,000 tracks per event, hit coordinate smearing enabled
	\item \texttt{exp-10k-events-10k-tracks} - 10\,000 events, 10\,000 tracks per event, hit coordinate smearing enabled
\end{itemize}

\subsection{Data set schema}
Considering that all of the above data sets are for the 3D domain, the schema and relevant elaborations for the generated data are listed below:
\begin{itemize}
	\item \texttt{event id} - An incremental identifier for events belonging to an experiment, which is unique within the scope of the experiment.
	\item \texttt{sub-detector id} - An incremental identifier for different sub-detector layers belonging to a geometry, which is unique within the scope of the geometry.
	\item \texttt{sub-detector type} - The type of the sub-detector layer recording a hit, which can be one of three available types, pixel, short-strip, or long-strip.
	\item \texttt{track id} - An incremental identifier for tracks belonging to an event, which is unique within the scope of the event.
	\item \texttt{track type} - Indicates the type of function defining the track in terms of polynomial degree. At the moment, all tracks are linear.
	\item $r_0$ or \texttt{radial const} - The $r$ coordinate of the $(r_0, \theta_0, z_0)$ tuple defining the point $P_0$, used in a track's parametric set of equations. The value will represent origin smearing for $r$. $r_0$ and \texttt{radial const} are applicable to \emph{linear} and \emph{helical expanding} track types, respectively.
	\item $\theta_0$ or \texttt{azimuthal const} - The $\theta$ coordinate of the $(r_0, \theta_0, z_0)$ tuple defining the point $P_0$, used in a track's parametric set of equations. The value will represent origin smearing for $\theta$. $\theta_0$ and \texttt{azimuthal const} are applicable to \emph{linear} and \emph{helical expanding} track types, respectively.
	\item $z_0$ or \texttt{pitch const} - The $z$ coordinate of the $(r_0, \theta_0, z_0)$ tuple defining the point $P_0$, used in a track's parametric set of equations. The value will represent origin smearing for $z$. $z_0$ and \texttt{pitch const} are applicable to \emph{linear} and \emph{helical expanding} track types, respectively.
	\item $r_d$ - The $r$ coordinate of the $(r_d, \theta_d, z_d)$ tuple defining the direction vector $V_d$, used in a track's parametric set of equations. $r_d$ is applicable to the \emph{linear} track type.\\
	OR,\\
	\texttt{radial coeff} - The coefficient affecting the radius rate in the helical track. \texttt{radial coeff} is applied to the free variable in the equation for $r$. \texttt{radial coeff} is applicable to the \emph{helical expanding} track type.
	\item $\theta_d$ - The $\theta$ coordinate of the $(r_d, \theta_d, z_d)$ tuple defining the direction vector $V_d$, used in a track's parametric set of equations. $r_d$ is applicable to the \emph{linear} track type.\\
	OR,\\
	\texttt{azimuthal coeff} - The coefficient affecting the clockwise/counter-clockwise extrusion direction in the helical track. \texttt{azimuthal coeff} is applied to the free variable in the equation for $\theta$. \texttt{azimuthal coeff} is applicable to the \emph{helical expanding} track type.
	\item $z_d$ - The $z$ coordinate of the $(r_d, \theta_d, z_d)$ tuple defining the direction vector $V_d$, used in a track's parametric set of equations. This value will be 1 or -1, depending on which side of the XY-plane the track is being extruded from. $r_d$ is applicable to the \emph{linear} track type.\\
	OR,\\
	\texttt{pitch coeff} - The coefficient affecting the pitch rate in the helical track. \texttt{pitch coeff} is applied to the free variable in the equation for $z$. \texttt{pitch coeff} is applicable to the \emph{helical expanding} track type.
	\item \texttt{hit id} - An incremental identifier for hits belonging to an event, which is unique within the scope of the event.
	\item $r_{hit}$ - The $r$ coordinate of the $(r_{hit}, \theta_{hit}, z_{hit})$ tuple defining the recorded hit point on the relevant sub-detector.
	\item $\theta_{hit}$ - The $\theta$ coordinate of the $(r_{hit}, \theta_{hit}, z_{hit})$ tuple defining the recorded hit point on the relevant sub-detector.
	\item $z_{hit}$ - The $z$ coordinate of the $(r_{hit}, \theta_{hit}, z_{hit})$ tuple defining the recorded hit point on the relevant sub-detector.
\end{itemize}

\subsection{Performance benchmarking}
In order to evaluate the performance of REDVID, we have benchmarked the execution of simulations with a lower event count, 1\,000 events per simulation and similar variations of track concentrations per event as before, i.e., $[1, 10\,000]$. For our metric collections, including CPU-time and execution duration, high-precision counters from the \texttt{time} library available in Python have been used. The collected CPU-time results are provided in \Cref{tab:execution_times}.
\begin{table*}[htbp]
	\centering
	\caption{REDVID execution CPU-time cost for simulations of 1\,000 events with various track concentrations. All values are in milliseconds. Full simulation times are provided in minutes as well. Even though REDVID is developed in Python, computational cost figures indicate efficiency for frequent executions.}
	\begin{tabular*}{\linewidth}{@{\extracolsep{\fill}}lrrrr@{}}
		\toprule
		\multicolumn{1}{c}{\textbf{\begin{tabular}[c]{@{}c@{}}Recipe for\\ 1\,000 events\end{tabular}}} & 
		\multicolumn{1}{c}{\textbf{3D detector spawning}} & 
		\multicolumn{1}{c}{\textbf{\begin{tabular}[c]{@{}c@{}}Track randomisation\\ per event - Mean\end{tabular}}} & 
		\multicolumn{1}{c}{\textbf{\begin{tabular}[c]{@{}c@{}}Hit discovery\\ per event - Mean\end{tabular}}} & 
		\multicolumn{1}{c}{\textbf{\begin{tabular}[c]{@{}c@{}}Full simulation of\\ 1\,000 events (minutes)\end{tabular}}}
		\\
		\midrule
		1 track per event      & 0.025                & 0.043                   & 1.463                 & 2\,731.17 (0.05)
		\\
		10 tracks per event    & 0.025                & 0.083                   & 13.429                & 15\,418.589 (0.26)
		\\
		100 tracks per event   & 0.025                & 0.465                   & 129.864               & 137\,623.954 (2.29)
		\\
		1\,000 tracks per event  & 0.025                & 4.582                   & 1\,285.989              & 1\,353\,396.641 (22.56)
		\\
		10\,000 tracks per event & 0.024                & 43.765                  & 12\,496.208             & 13\,591\,628.526 (226.53)
		\\
		\bottomrule
	\end{tabular*}
	\label{tab:execution_times}
\end{table*}

Simulations have been performed on the DAS-6 compute cluster~\cite{Bal:2016:MDSC}. The machines used are each equipped with a single 24-core AMD EPYC 7402P processor and 128~GB of main memory. Note that the mean CPU-time calculations do not include the first event of each recipe batch. This is due to the presence of cold-start effect for the first event and delays resulting from it.

Though we have enforced single-threaded operation for our benchmarks, workload parallelisation is rather trivial. The number of events to be generated can be divided into any desired number of batches and distributed amongst multiple threads. Considering the timing results, we observe that the CPU-time values scale linearly, i.e., a tenfold increase in the track concentration per event results in roughly a tenfold increase in the full simulation CPU-time.

\subsection{User operation}
Whether independently, or as an integrated module within a workflow, similar to the depiction from \Cref{fig:simulation_application}, users can use REDVID primarily to generate data sets. The main script to execute the tool is \texttt{digital\_detector.py}~\cite{Odyurt:2023:REDVIDREPO}. A configuration file is included and populated with parameter values. Users only have to change the \texttt{anchor\_path} parameter to a valid path. This will be system dependent. Alternatively, a different configuration file path can be provided as an argument. The default name is \texttt{REDVID\_config.ini}, which can also be changed. Python package dependencies are minimal and can be observed in the \texttt{requirements.txt} file.

\section{Related work}
\label{sec:related_work}
Although the overall available data is abundant, corner case data is rather scarce. Such real-world data, or data synthesised with accurate (in our case physics-accurate) simulations is complex in terms of data dimensionality and granularity. This complexity is directly resulting from the complexity of the real system, or the accurate (physics-accurate) model of the system in case of simulations. Within the HEP landscape for instance, we touched upon the complexity of simulators such as Geant4 in \Cref{sec:background}, as well as the dependence on these simulators by tools like ATLFAST.

The first challenge, lack of annotated data for one or more specific scenarios, has been recognised in the literature~\cite{deMelo:2022:NGDL}. The second challenge though, the issue of complexity, is not as well known. A closely related acknowledgement has been made regarding the complexity level of models for simulations~\cite{Chwif:2000:SMC}.

The two main shortcomings of the previous efforts towards the use of ML in physics problems have been use-case specificity~\cite{Willard:2022:ISKM} and the lack of user-friendly tools~\cite{Carleo:2019:MLPS}. As noted by Willard et al.~\cite{Willard:2022:ISKM}, the efforts surrounding the use of ML for physics-specific problems are focused on sub-topics, or even use-cases. Although our methodology and synthetic data focuses on the domain of tracking for detector data, we could claim that  it is independent of the chosen detector experiment.

The point from~\cite{Willard:2022:ISKM} regarding the computational efficiency of ROMs matches our motivation. Where our work differs is in the placement of our ROM within our methodology. Our reduced model of a detector is considered as the model for simulations resulting in synthetic data generation, which is different than ML-based surrogate models as ROMs~\cite{Chen:2012:SVMR, Kasim:2022:BHAE}, or ML-based surrogate models built from ROMs~\cite{Xiao:2019:ROMT}.

\section{Conclusion and future work}
\label{sec:conclusion}
With many computational science applications exploring the use of ML-assisted solutions, there is a need for reduced complexity simulations to facilitate the design process. In this work, we show how a reduction in simulation complexity through ROMs and a smaller behavioural-space for the simulator can result in a lower complexity for synthesised data. This is particularly relevant for our HEP use-case.

To demonstrate the feasibility of this approach, we have presented the design and implementation details of REDVID (REDuced VIrtual Detector), our simulation framework fulfilling such a reduction. To demonstrate REDVID's feasibility, we executed it with relevant workload recipes, and have made available the resulting data sets over Zenodo open repository. We further analysed the computational cost figures for these experiments, and we conclude that, even though our tool is developed in Python, computational cost figures (case in point, 15 seconds, 138 seconds and 22 minutes of CPU-time for 1\,000 events with 10, 100 and 1\,000 tracks per event, respectively) indicate efficiency for frequent executions. Accordingly, the lightweight nature of REDVID simulations makes our tool a suitable choice as a simulation-in-the-loop with data-driven workflows for HEP. This is the case of searching for a ML-assisted solution to address the challenge of particle track reconstruction.

However, reduced complexity and less descriptive data distances our simulations from the physics-accurate ground truth. We have explained that, to opt for such an approximation, is a deliberate act, positioning REDVID as a suitable middle ground amongst other available tools, not as exact as physics-accurate simulations, and not as synthetic as dummy data generators. The reduced complexity especially allows for early problem formulation and testing at early stages, when dealing with ML-assisted solution design workflows. Yet another advantage of reduced complexity data that still respects the high-level relations, is in its pedagogical merit, enabling problem solving practices in higher education.

While keeping the distance from physics-accurate tools, REDVID can be extended in numerous ways. Considering our foreseen methodology, we will be implementing further low-cost, complexity inducing features, e.g., various track randomisation protocols to allow for diverse particle propagation scenarios, more complex non-linear track definitions beyond helical tracks, and possibly a Domain-Specific Language (DSL) to be used for virtual detector definitions.

Aside from REDVID itself, we intend to implement the full ML-assisted solution search workflow depicted in \Cref{fig:simulation_application} and perform explorations of models based on different ML architectures.

\begin{acks}
This project is supported by the Nederlandse Organisatie voor Wetenschappelijk Onderzoek (NWO), a.k.a., the Dutch Research Council (grant no. 62004546).
\end{acks}


\balance

\printbibliography

\end{document}